\documentclass[pra,preprint,aps,showpacs,floatfix,superscriptaddress]{revtex4}

\usepackage{graphicx}
\usepackage{rotating}
\usepackage{amsmath}
\usepackage{bbm}
\usepackage{color}

\renewcommand{\v}[1]{{\bf #1}}

\newcommand{\be}{\begin{equation}}
\newcommand{\ee}{\end{equation}}
\newcommand{\bea}{\begin{eqnarray}}
\newcommand{\eea}{\end{eqnarray}}

\begin{document}

\title{Exact Kohn-Sham potential of strongly correlated finite systems}
\author{N. Helbig}
\affiliation{Nano-Bio Spectroscopy group and ETSF Scientific Development Centre, 
Dpto. F\'isica de Materiales, Universidad del Pa\'is Vasco, Centro de
F\'isica de Materiales CSIC-UPV/EHU-MPC and DIPC, Av. Tolosa 72, E-20018 San 
Sebasti\'an, Spain}
\author{I.V. Tokatly}
\affiliation{Nano-Bio Spectroscopy group and ETSF Scientific Development Centre, 
Dpto. F\'isica de Materiales, Universidad del Pa\'is Vasco, Centro de
F\'isica de Materiales CSIC-UPV/EHU-MPC and DIPC, Av. Tolosa 72, E-20018 San 
Sebasti\'an, Spain}
\affiliation{IKERBASQUE, Basque Foundation for Science, 48011, Bilbao, Spain}
\author{A. Rubio}
\affiliation{Nano-Bio Spectroscopy group and ETSF Scientific Development Centre, 
Dpto. F\'isica de Materiales, Universidad del Pa\'is Vasco, Centro de
F\'isica de Materiales CSIC-UPV/EHU-MPC and DIPC, Av. Tolosa 72, E-20018 San 
Sebasti\'an, Spain}
\affiliation{Fritz-Haber-Institut der Max-Planck-Gesellschaft, Berlin, Germany}

\begin{abstract}
The dissociation of molecules, even the most simple hydrogen molecule, cannot be
described accurately within density functional theory because none of the 
currently available functionals accounts for strong on-site correlation. This
problem has led to a discussion of properties that the local Kohn-Sham potential
has to satisfy in order to correctly describe strongly correlated systems. We
derive an analytic expression for this potential at the dissociation limit and
show that the numerical calculations for a one-dimensional two electron model
system indeed approach and reach this limit. It is shown that the functional
form of the potential is universal, i.e. independent of the details of the
system.
\end{abstract}

\date{\today}

\maketitle

\section{Introduction}\label{sec_intro} 

Over the years the improvement in exchange-correlation (xc) functionals has
made  density functional theory (DFT)\cite{HK1964,KS1965} the tool of choice to
accurately study and predict properties of many-electron systems. Applications
range from atoms to molecules and nanostructures, biomolecules and solids and
cover diverse topics such as theoretical spectroscopy, e.g. optical, energy
loss, and time-resolved spectroscopy, electron transport, light induced phase
transitions, photochemistry, and electrochemistry \cite{MR2009,CCJ2009}. Despite
this success major basic challenges remain that usually are manifestations of
strong, static and dynamic, electron correlations \cite{FNGB2005}. Van der Waals
interactions, the localization in strongly-correlated systems, open-shell
molecules, and molecular dissociation are poorly accounted for by present
functionals \cite{FNGB2005,D2006}. A general measure of inter-electron
correlations is the ratio of the kinetic energy to the potential energy of the
Coulomb interaction between electrons. While the kinetic energy is lowered by
delocalization of electrons over the system, the Coulomb repulsion works in the
opposite direction trying to keep electrons far from each other and thus
favoring the tendency to localization. In the condensed matter context this
interplay of two opposite tendencies is commonly pictured in terms of the
Hubbard on-site correlations that suppress tunneling of particles between atoms
and lead to localization of electrons on lattice sites (or groups of sites).
Strong Hubbard correlations are responsible for the dissociation of molecules,
the physics of Mott insulators, non-itinerant magnetism in most of the magnetic
dielectrics, the Coulomb blockade in quantum transport, etc. The failure of the
common DFT-functionals to capture the effects of Hubbard correlations led to the
development of the LDA+U method \cite{AAL1997} and its more elaborated
counterpart, the dynamical mean-field theory (DMFT) \cite{DMFT}, to describe
strongly correlated systems. On the other hand, it is absolutely clear that DFT
being an ``in principle exact'' theory should be capable to describe the regime
of strong correlations provided the proper xc potential is known. In this realm,
it is fundamental to increase the knowledge of relations, fulfilled by the exact
xc potential, in order to move forward on the road towards the ultimate
functional, the ``holy-grail of DFT''. 

In the present work, we consider a prototypical example of a physical behavior
governed by strong Hubbard correlations -- the dissociation of diatomic
molecules, and discuss exact features of the xc potential $v_{xc}$ necessary to
describe the correlation-driven electron localization happening in the
dissociation limit.  One such feature is well known -- in the dissociation of
heteroatomic molecules the Kohn-Sham (KS) potential $v_s$ acquires a step in
between the fragments to adjust the ionization potentials \cite{AB1985}. The
value of this step is universal and is simply given by the difference of the
ionization potentials of the two fragments of the dissociated molecule.
Apparently, the presence of this step is necessary to prevent an unphysical fall
of electrons to the fragment with a higher ionization potential. However, as we
show below, it is not sufficient to correctly describe the dissociation, i.e.
the strongly correlated, limit. In fact, in this limit the xc potential acquires
a nontrivial structure even for the most simple homoatomic molecules, such as
H$_2$. 

An important step in understanding the behavior of the xc potential in the
dissociation limit has been made in a series of works by Baerends and co-authors
\cite{BBS1989,LB1994,GRB1995,GB1996,GB1997} who reconstructed the xc potential
of a number of stretched diatomic molecules from an accurate many-body
configuration interaction (CI) ground-state wave function. They noticed
numerically that, in addition to the step, $v_{xc}$ also shows a peak structure
around the middle point between the two atoms
\cite{BBS1989,LB1994,GRB1995,GB1996}. A subsequent analysis has shown that the
peak in $v_{xc}$ is probably a general feature of the dissociation limit, which
contradicts the common LCAO form of the molecular orbital, but can be reasonably
well reproduced assuming that the two-electron wave function is of
Heitler-London form constructed out of the atomic KS orbitals
\cite{GB1997,TMM2009}. The physical nature of this peak structure and its
connection to the Hubbard correlations is the main subject of our paper. We
prove that the whole spatial dependence of the KS potential in the strongly
correlated dissociation limit, including both the peak structure and the step
(for heteroatomic molecules), is universal. It depends only on the asymptotic
behavior of the density of the fragments, which, in turn, is mainly determined
by the atomic ionization potentials \cite{AB1985}. In particular, we derive an
analytic formula that allows us to recover the {\em exact} form of the xc
potential in the dissociation limit from the knowledge of the ionization
potentials of the independent fragments. This result adds one more item to the
list of exact properties of the KS system and xc potential, such as Koopman's
theorem, the exact asymptotic form of $v_{xc}$ for finite systems, and the exact
relation of the asymptotics of the density to the asymptotics of the highest
occupied KS state \cite{AB1985}. We also demonstrate that the peak structure in
$v_{xc}$ can be viewed as a manifestation of the Hubbard on-site correlations at
the level of noninteracting KS particles. The physical significance of this peak
is that it suppresses the quantum tunneling of KS particles between two
fragments, exactly what the Hubbard repulsion does for real electrons. This
ensures that the fragments become physically independent. We emphasize that this
effect is not accounted for by any of the currently available functionals and
constitutes a stringent test for the future development of static and
time-dependent functionals aimed at describing strongly correlated systems.

The structure of the paper is as follows. In Sec.~\ref{sec_disslimit} we discuss
the physics of the strongly correlated dissociation limit in terms of both
Hubbard on-site correlations and the KS formulation of DFT. Using a simple
analytically solvable model for a 1-dimensional (1D) symmetric diatomic we
derive the asymptotic form of the KS potential and verify our findings
numerically for more general symmetric 1D systems. In Sec.~\ref{sec_kspot} we
uncover the universal physics that governs the behavior of the KS potential in
the dissociation limit, derive general exact analytic formulas valid for all
two-electron systems and verify them numerically for model 1D heteroatomic
molecules. We also discuss generalizations of the results for more general
many-electron systems. We then conclude the paper by summarizing our main
results.  

\section{Towards DFT in the dissociation limit}\label{sec_disslimit}

\subsection{Physics of the dissociation limit: Real electrons vs. Kohn-Sham particles}

Let us first consider the qualitative physics of the dissociation of simple
diatomic molecules with a single $\sigma$-bond formed by a pair of electrons
originating from the atomic valence orbitals. Specific examples for this
scenario include H$_2$, Li$_2$, and LiH, to name a few. When the molecule is
stretched the gain in the kinetic energy due to the delocalization of the
electrons, which is proportional to the hopping matrix element $t$, decreases
exponentially. On the other hand, the loss in the interaction energy, due to the
presence of two electrons on the same atom, saturates at a certain value of the
Hubbard on-site repulsion $U$. Starting from some distance, roughly determined
by the condition $U\gtrsim t$, the on-cite Coulomb correlations block the
inter-atomic tunneling, the electrons get localized on their own atoms, and the
molecule dissociates into two physically independent fragments. 

Within DFT the real interacting system is modeled by an artificial
non-interacting KS system with the same ground-state density. The
non-interacting particles are subject to an effective potential via the KS
\cite{KS1965} equation (atomic units are used throughout the paper)
\be\label{kseq}
\left[-\frac{\nabla^2}{2}+v_s(\v r)\right]\varphi_j(\v r)=
\epsilon_j\varphi_j(\v r).
\ee
Since the KS particles are noninteracting there is no way to localize them on a
particular atom, independent of the distance $d$ between the fragments. For a
symmetric molecule, like H$_2$ or Li$_2$, the KS particles responsible for the
formation of the bond always occupy a symmetric orbital with a probability of
$1/2$ to find either particle on each atom. 

Apparently, the behavior of the KS particles is very different from that of real
physical electrons. The difference between the real word and an artificial word
of KS particles becomes especially striking in the regime of strong
correlations, and the dissociation of simple molecules provides us with a bright
example of this phenomenon. However, a certain physical information, namely the
ground-state density, is reproduced exactly by the KS system. Therefore, the
real physics should be reflected in the properties of the KS system.
Establishing a map of the physics governed by the strong Hubbard on-site
correlations to the properties of the KS potential, i.e. the map of the real
word to the world of KS particles, is the main subject of this work. In order to
find this map we mainly concentrate on a minimal model that captures all key
physics of dissociation -- the system of two electrons in a potential formed by
two nuclei/potential wells. In Sec.~\ref{sec_kspot} we argue that the main
conclusions are transferable to a more general many-electron case.

In the case of two electrons in a singlet state only one spatial KS orbital is
occupied. Therefore, the density is given as $n(\v r)=2|\varphi_1(\v
r)|^2=2\varphi_1^2(\v r)$, because the orbital can always be chosen to be
real. Hence, from inverting Eq. (\ref{kseq}), the exact KS potential is given by
\be\label{KSpot}
v_s(\v r)=\frac{1}{2}\frac{\nabla^2\sqrt{n(\v r)}}{\sqrt{n(\v r)}} 
+\epsilon_1
\ee
with $n(\v r)$ being, by construction, the exact ground-state density of the
two-electron system. Hence, given the exact two-body ground-state wave function
$\Psi(\v r_1, \v r_2)$ one can calculate the density $n(\v r) = \int d\v
r_2|\Psi(\v r, \v r_2)|^2$, and then recover the exact KS potential by inserting
$n(\v r)$ into Eq.~(\ref{KSpot}). This formally maps the physical two-body wave
function to the KS potential. However, extracting the physics behind this formal
map is not as simple as one may think since in a general 3D case the wave
function $\Psi(\v r_1, \v r_2)$ is a complicated object given fully numerically,
e.g. from CI calculations, and, moreover, may be numerically problematic for
realistic systems when one reaches the dissociation limit . Therefore, it is
instructive to look first at some simplified models and then, after the
essential physics is understood, return to realistic situations.

An obvious simplification, which still contains all physical ingredients of the
original problem, is to consider a system of two interacting particles in one
dimension. The corresponding two-electron Schr\"odinger equation takes the form
\begin{equation}
 \label{2El-SE}
\left[-\frac{1}{2}\left(\frac{\partial^2}{\partial x_1^2} + 
\frac{\partial^2}{\partial x_2^2}\right) +
v_{\mathrm{ext}}(x_1) + v_{\mathrm{ext}}(x_2) + 
v_{\mathrm{int}}(|x_1-x_2|)\right]\Psi(x_1,x_2)=E\Psi(x_1,x_2),
\end{equation}
where $v_{\mathrm{ext}}(x)$ is the external potential, and 
$v_{\mathrm{int}}(|x-x'|)$ is the potential of the inter-particle interaction. At
the end of this section and in Sec.~\ref{sec_kspot} we present the results based
on the full numerical solution of Eq.~(\ref{2El-SE}). However, to gain some 
physical insight into the shape of $v_s$ in the dissociation limit we simplify
the model even further to make it analytically solvable.

\subsection{Analytical model of strongly correlated electrons}

First, we assume that the external potential in Eq.~(\ref{2El-SE}) is given by a
sum of two $\delta$-function wells of equal strength, $v$, located at the
points $x=\pm d/2$. Similarly, we take the interaction to be a zero-range
delta-potential of strength $\lambda$
\begin{eqnarray}
\label{Vext-delta}
v_{\mathrm{ext}}(x) &=& -v[\delta(x-d/2)+\delta(x+d/2)],\\
\label{Vint-delta}
v_{\mathrm{int}}(|x-x'|) &=& \lambda \delta(x-x').
\end{eqnarray}
Physically, in the dissociation limit the only role of the interaction is to
block the inter-atomic tunneling. Therefore, in that limit, the behavior is
expected to be universal and independent of a particular form and/or strength of
the interaction. This leads us to the last simplifying assumption, namely the
limit of infinitely strong $\delta$-repulsion, $\lambda\to\infty$. Now the
problem becomes immediately solvable by the so called Girardeau mapping
\cite{Girardeau1960} (see also a more recent review \cite{YukGir2005}), which
allows to map the ground state of strongly interacting ``hard-core'' bosons (a
symmetric wave function) to the ground state of noninteracting fermions
(antisymmetric wave function). In our two-particle case the exact ground-state
(singlet, i.e. symmetric) wave function takes the form
\begin{equation}
 \label{GS-delta}
\Psi(x_1,x_2) = |\phi_1(x_1)\phi_2(x_2) - \phi_2(x_1)\phi_1(x_2)|,
\end{equation}
where $\phi_1(x)$ and $\phi_2(x)$ are the two lowest states of the following 
one-particle Schr\"odinger equation
\begin{equation}
\label{Gir-SE}
-\frac{1}{2}\phi''_n(x) + v_{\mathrm{ext}}(x)\phi_n(x)=\epsilon_n\phi_n(x).
\end{equation}
In other words, the ground state of two infinitely interacting particles in a
singlet state is given by the modulus of the ground state wave function of two
noninteracting spinless fermions in the bare external potential
$v_{\mathrm{ext}}(x)$.

The two lowest energy solutions of Eq.~(\ref{Gir-SE}) with the external
potential of Eq.~(\ref{Vext-delta}) are easily found to be
$\phi_1(x)=\phi_{+}(x)$ and $\phi_2(x)=\phi_{-}(x)$ with
\begin{equation}
\label{Psi-pm}
\phi_{\pm}(x)=C_{\pm}\left(e^{-\alpha_{\pm}|x+d/2|} 
\pm e^{-\alpha_{\pm}|x-d/2|}\right),
\end{equation}
where $C_{\pm}$ are the normalization constants. The parameters $\alpha_{\pm}$,
which determine the corresponding eigenvalues
$\epsilon_{1,2}=\epsilon_{\pm}=-\alpha_{\pm}^2/2$, are the solutions of the
following dispersion equations \footnote{We only consider distances $d>1/v$
where both lowest states are bound.} 
\begin{equation}
 \label{alpha-pm}
\alpha_{\pm} = v\left(1 \pm e^{-\alpha_{\pm}d}\right).
\end{equation}
Using the exact ground-state wave function (\ref{GS-delta}) we obtain the exact
density
\begin{equation}
\label{n-delta}
n(x) = \int dx' \Psi^2(x,x') = \phi_{+}^2(x) + \phi_{-}^2(x),
\end{equation}
and, finally, by inserting $n(x)$ into the 1D version of Eq.~(\ref{KSpot}), the
exact KS potential for our strongly correlated two-particle system, 
$v_s(x)=\Delta v_s(x) + v_{\mathrm{ext}}(x)$,
\begin{equation}
 \label{Vs1-delta}
\Delta v_s(x) = \frac{(\phi_{+}\phi'_{-} - \phi'_{+}\phi_{-})^2}
{2(\phi_{+}^2 + \phi_{-}^2)^2} -
\frac{\epsilon_{+}\phi_{+}^2 + \epsilon_{-}\phi_{-}^2}
{\phi_{+}^2 + \phi_{-}^2} - \frac{v^2}{2}.
\end{equation}
Equation~(\ref{Vs1-delta}) gives the exact KS potential for any distance between
the wells. In the dissociation limit, $v d\gg 1$, $\alpha_{\pm}\to v$ and
$\epsilon_{\pm}\to -v^2/2$. Therefore, the last two terms in
Eq.~(\ref{Vs1-delta}) cancel while the remaining first term simplifies to
\begin{equation}
 \label{Vs-delta}
\Delta v_s(x) = \frac{v^2}{2\cosh^2(2v x)} \equiv 
\frac{I}{\cosh^2(2\sqrt{2I} x)}.
\end{equation}
$I=v^2/2$ is the ionization potential of a separate fragment, the 
delta-potential of strength $v$. Hence, we have found that the exact KS 
potential in the dissociation limit has the form of a ``wall'' built up between 
the two fragments of the ``molecule''.  The shape of this wall looks quite 
close to the peak structure observed numerically in previous works 
\cite{BBS1989,LB1994,GRB1995,GB1996}. 

\subsection{1D model for homoatomic dissociation}

It is physically plausible to expect that the behavior in the dissociation limit
is independent of the particular form and strength of the interaction, and that
the asymptotic form of $v_s(x)$ for more general systems is similar to that
given by the simple formula (\ref{Vs-delta}). We now verify this expectation for
a 1D system of two particles in a more general, but still symmetric, external
potential, namely
\be\label{Vext-sym}
v_{\mathrm{ext}}(x)=-v
\left[\frac{1}{\cosh^2(x-d/2)}+\frac{1}{\cosh^2(x+d/2)}\right].
\ee
The two particles are interacting with a finite range interaction potential of 
the form
\be\label{Vint}
v_{\mathrm{int}}(|x-y|)=\frac{b}{\cosh^2(x-y)}.
\ee
The choice of the $1/\cosh^2$ shape of the wells and the interaction potential
is arbitrary. It is simply a matter of convenience as the 1D Schr\"odinger
equation with a $1/\cosh^2$ potential is exactly solvable \cite{LL1977}, which
allows us to control the accuracy of our numerical calculations. In addition,
the finite-range interaction (\ref{Vint}) allows us to reach the dissociation
limit in a controllable way without numerical instabilities. 

For the numerical solution of Eq.~(\ref{2El-SE}) with general $v_{\mathrm{ext}}$ 
and $v_{\mathrm{int}}$, we note that the 1D two-particle problem defined by 
Eq.~(\ref{2El-SE}) can be formally interpreted as a 2D one-particle problem 
with the Hamiltonian
\be\label{2Dham}
H^{\mathrm{2D}} = -\frac{1}{2}\left[\frac{\partial^2}{\partial x^2}+\frac{\partial^2}{\partial y^2}\right]
+ v_{\mathrm{ext}}^{\mathrm{2D}}(x,y),
\ee
where the effective 2D one-particle potential is defined as
\be
v_{\mathrm{ext}}^{\mathrm{2D}}(x,y)=
v_{\mathrm{ext}}(x)+v_{\mathrm{ext}}(y)+v_{\mathrm{int}}(|x-y|).
\ee
Consequently, the exact ground-state wave function $\Psi (x,y)$ and the exact
one-dimensional ground-state density for the physical two-particle system,
$n(x)=\int dy |\Psi (x,y)|^2$, can be obtained numerically from any computer
code that is able to treat non-interacting electrons in two dimensions. All our
calculations in this work were carried out with the \texttt{OCTOPUS} code
\cite{MCBR2003}. 

\begin{figure}
\includegraphics[width=0.48\textwidth,clip]{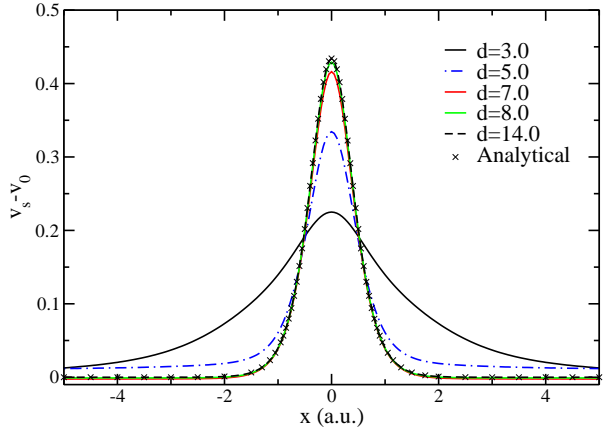}
\caption{\label{fig:symmwall}Kohn-Sham potential for two equivalent wells at
different distances ($v=0.9$). The external potential has been subtracted to facilitate the
comparison. Analytical results are given by Eq. (\ref{Vs-delta}).}
\end{figure}

The exact KS potential, $v_s(x)$, for $v_{\mathrm{ext}}$ and $v_{\mathrm{int}}$ 
of Eqs.~(\ref{Vext-sym}) and (\ref{Vint}) with $v=0.9$ and $b=0.5$, and varying 
interwell distance $d$ is shown in Fig.~\ref{fig:symmwall}. At first sight the
results look very surprising: starting from a certain distance, $d=8$~a.u. for
these particular parameters, the shape of the KS potential saturates exactly at
the form given by the analytic formula (\ref{Vs-delta}) with $I$ being the
ionization potential of a single $1/\cosh^2$ well. Calculations for different
strengths of the wells $v$, different interaction strength $b$, as well as for a
long-range soft-Coulomb inter-particle interaction all show the same result
\footnote{We note that in the case of a long-range soft Coulomb interaction we
have encountered numerical instabilities at too large distances.}. At large
distances the exact KS potential is not only similar to the analytic form of
Eq.~(\ref{Vs-delta}), obtained from an oversimplified model with an infinite
$\delta$-repulsion, but matches it {\em exactly} as soon the dissociation limit
is reached! In the next section we discuss a deep, though simple physical reason
for this seemingly surprising universality.

\section{Exact Kohn-Sham potential in the dissociation limit}\label{sec_kspot}

\subsection{Universality of the Kohn-Sham potential}
In order to understand the nature of the universal peak in the asymptotic form
of the KS potential we turn back to our first simple model with an infinite
zero-range repulsion and look more closely at the behavior of the exact density
determined by Eq.~(\ref{n-delta}). In the dissociation limit, $v d\gg 1$, the
functions $\phi_{+}(x)$ and $\phi_{-}(x)$ become simple symmetric and
antisymmetric combinations of ``atomic'' orbitals. Taking the squares and
summing them up, as suggested by Eq.~(\ref{n-delta}), we find that all
interference terms, i.e. the cross-product of different atomic orbitals, cancel, 
and the total density reduces to a sum of two atomic densities
\begin{equation}
\label{n2-delta}
n(x) = v e^{-2v |x-d/2|} + v e^{-2v |x+d/2|}.
\end{equation}
This is exactly what Hubbard on-site correlations do -- they destroy the
inter-atomic tunneling/interference, which localizes the electrons on separate
sites, and eventually makes the density to be the sum of the densities of two
physically independent fragments. On the KS side of the mirror, the KS
potential, whatever it is, cannot localize the KS particles. However, by
building up a self-consistent wall between the fragments it suppresses the
tunneling/interference of the atomic KS orbitals to mimic the density
distribution of the two independent atoms. Thus, the physics of the on-site
Hubbard correlations in the real world is mapped to the wall in the KS potential
in the artificial world of KS particles. It is, therefore, not surprising that
the universality of the physics in the dissociation limit is reflected in the
universal form of the asymptotic KS potential. Since for DFT only the density
distribution is essential, the general condition that determines the KS
potential in the dissociation limit is simply
\begin{equation}
\label{n-diss}
n(\v r) = n_1(\v r) + n_2(\v r),
\end{equation}
in other words, the total density $n(\v r)$ is equal to the plain sum of the 
densities, $n_1(\v r)$ and $n_2(\v r)$, of the two independent fragments. The 
asymptotic form of the KS potential should be such that it supports the density 
distribution given by Eq.~(\ref{n-diss}). As the densities $n_1(\v r)$ and 
$n_2(\v r)$ decay exponentially from different sides the only way to mimic this 
at the level of a single KS orbital is to insert a potential wall in the middle 
region.

Having understood the key physics we are ready to go to more complex systems.

\subsection{Kohn-Sham potential of heteroatomic 1D molecules}

It is now straightforward to find the form of the KS potential in the
dissociation limit for a general 1D molecule formed by two different wells.
Assuming that the densities, $n_1(x)$ and $n_2(x)$, corresponding to one electron
sitting in a separate well are known, we require that the total density $n(x)$
is given by their sum, Eq.~(\ref{n-diss}), and substitute this sum into
Eq.~(\ref{KSpot}). The result can be reduced to a form that looks structurally
similar to Eq.~(\ref{Vs1-delta})
\begin{equation}
 \label{Vs1}
\Delta v_s(x) = \frac{\left[\sqrt{n_1}\partial_x\sqrt{n_2} - \sqrt{n_2}\partial_x\sqrt{n_1}\right]^2}{2(n_1 + n_2)^2} +
\frac{I_1n_1 + I_2n_2}{n_1 + n_2} - I,
\end{equation}
where $I_{1,2}$ are the ionization potentials of the fragments and 
$I=\min\{I_1,I_2\}$ is the ionization potential of the total system. 
Equation~(\ref{Vs1}) is valid in the dissociation limit, and from its structure 
it is clear that $\Delta v_s(x)$ has a nontrivial $x$-dependence (i.e. differs 
from a constant) only far away from the ``atoms'', where the densities fall off 
exponentially. Therefore, for the practical evaluation of $v_s$ in the 
dissociation limit, it is sufficient to know only the asymptotic behavior of the 
density of the separate fragments. In the 1D case, the asymptotics of the 
densities  $n_1(x)$ and $n_2(x)$ have the following general form
\begin{equation}
\label{1d-n}
n_{1,2}(x) = A_{1,2}e^{-2\alpha_{1,2}|x\pm d/2|},
\end{equation}
where the exponents $\alpha_{1,2}$ are related to the ionization potentials of 
the atoms $I_{1,2}=\alpha_{1,2}^2/2$ and $A_{1,2}$ are prefactors to the
exponential decay. 

Inserting Eq.~(\ref{1d-n}) for a symmetric molecule (equivalent wells with 
$A_1=A_2$, and $\alpha_1=\alpha_2$) into Eq.~(\ref{Vs1}) we immediately recover 
our first model result of Eq.~(\ref{Vs-delta}) thus confirming its
universality.  In a general asymmetric case (different wells or a
``heteroatomic'' molecule) a new qualitative feature, a ``shelf'', appears.
This shelf in $v_s$ is such that it aligns the ionization potentials of the
atoms. Formally, it results from the last two terms in Eq.~(\ref{Vs1}) which do
not cancel if the ionization potentials are different. Substituting the general
form of Eq.~(\ref{1d-n}) into Eq.~(\ref{Vs1}) we find the explicit results for
$v_s$ in different regions of space. It is convenient to represent the KS
potential as a sum of two different contributions, i.e. $\Delta
v_s(x)=v_s^{(1)}(x)+v_s^{(2)}(x)$. For the region between the wells, i.e. for
$-d/2<x<d/2$, the two contributions correspond to the ``wall'' and the ``shelf''
discussed before. They are given by
\bea\label{vspeak}
v_s^{(1)}(x)&=&
\frac{(\alpha_1+\alpha_2)^2/8}{\cosh^2[(\alpha_1+\alpha_2)(x+x_0)]},\\
\label{vsshelf1}
v_s^{(2)}(x)&=&
\frac{I_2-I_1}{1+\exp[2(\alpha_1+\alpha_2)(x+x_0)]}
\eea
with
\be
x_0=\frac{1}{\alpha_1+\alpha_2}\left[\frac{\alpha_1-\alpha_2}{2}\cdot
d+\log\frac{A_2}{A_1}\right].
\ee
Here, and in the following, we have assumed that $\alpha_1\leq\alpha_2$, i.e.
that the left fragment has a larger ionization potential. Obviously,  
for a symmetric configuration, $v_s^{(2)}$ vanishes identically, i.e. there
is only a peak in this case. Also, in this case $x_0=0$, i.e. the peak is
exactly in the middle between the two identical fragments as expected from
symmetry. 

For $x<-d/2$ the two contributions read
\bea
v_s^{(1)}(x)&=&
\frac{(\alpha_1-\alpha_2)^2/8}{\cosh^2[(\alpha_1-\alpha_2)(x+x'_0)]},\\
v_s^{(2)}(x)&=&
\frac{I_2-I_1}{1+\exp[-2(\alpha_1-\alpha_2)(x+x'_0)]}
\eea
with
\be\label{xzeroprime}
x'_0=\frac{1}{\alpha_1-\alpha_2}\left[\frac{\alpha_1+\alpha_2}{2}\cdot
d-\log\frac{A_2}{A_1}\right].
\ee
Contrary to before, $v_s^{(1)}$ does not describe a peak but it can actually be
shown that the potential is strictly monotonically increasing describing the
building up of the shelf, or its return to zero depending on the direction
one approaches $x'_0$ from. Also, for the symmetric case, both contributions
vanish as there is no shelf in that case. For  the region $x>d/2$ the potential
decays exponentially without specific features. 

We emphasize that neither the specific form of the fragments nor  the type of
interaction between the electrons enters the derivation of our analytical result
directly.  The specifics of the fragments appear in the result only via the
parameters $\alpha_{1,2}$ and $A_{1,2}$. The former describes how fast the
density decays, i.e. it is directly related to the ionization potential of each
fragment. The latter is connected to the normalization of the wave function and
only enters the potential as a logarithmic correction to the position of the
peak and shelf without changing the shape of the potential. For a symmetric
system the potential is completely determined by the ionization potential of the
two fragments. In all cases, symmetric and asymmetric, the functional form of
the KS potential is universal, only the position and the width and height of the
peak depend on the system under consideration. Both in the symmetric and in the
asymmetric case the presence of the universal wall reflects Hubbard
correlations. The potential wall suppresses the tunneling and drives the KS
density to the density corresponding to physically independent subsystems.

To ensure that our universal analytical formulas are indeed correct we performed
numerical calculations for an asymmetric two-electron system with the external
potential given by the sum of two different potential wells
\be\label{wells}
v_{\mathrm{ext}}(x)=-\frac{v_1}{\cosh^2(x-d/2)}-\frac{v_2}{\cosh^2(x+d/2)}
\ee
with $v_1=0.9$ and $v_2=0.7$. As before, for the interaction we keep the finite 
range potential of Eq.~(\ref{Vint}). 
Since the one-particle problem with $1/\cosh^2$ is exactly solvable
\cite{LL1977} the parameters $\alpha_{1,2}$ and $A_{1,2}$, entering our
asymptotic formulas Eqs.~(\ref{vspeak})-(\ref{xzeroprime}), are available in the
analytic form. In particular, for the pre-exponential factors in the asymptotics 
of the ``atomic'' densities we get
\be
A_{1,2}=2^{2\alpha_{1,2}}
\frac{\Gamma(\alpha_{1,2}+1/2)}{\sqrt{\pi}\Gamma(\alpha_{1,2})},
\ee
where $\Gamma$ denotes the usual Gamma-function. 

\begin{figure}
\includegraphics[width=0.48\textwidth,clip]{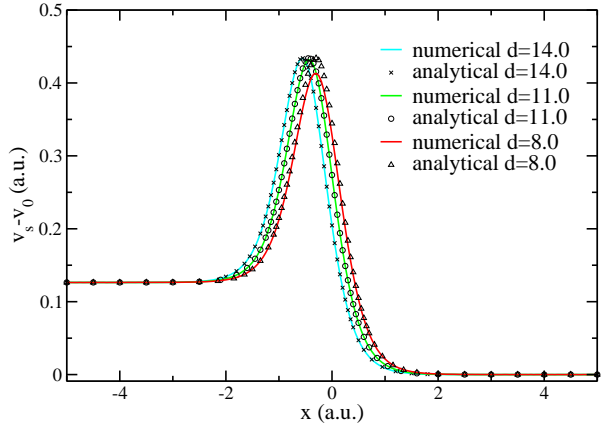}
\caption{\label{fig:asymmwall}Kohn-Sham potential for two different wells at
different distances ($v_1=0.9$, $v_2=0.7$). The external potential has been subtracted to facilitate the
comparison. Analytical results are given by Eqs. (\ref{vspeak}) and
(\ref{vsshelf1}). The potential returns to zero at large negative $x$.}
\end{figure}

In Fig.~\ref{fig:asymmwall} we show the comparison of the analytic KS potential
given by Eqs. (\ref{vspeak}) and (\ref{vsshelf1}) and the KS potential obtained
from the full numerical solution of the problem defined by Eqs.~(\ref{2El-SE}),
(\ref{Vint}), and (\ref{wells}). As expected, in the asymmetric case, $v_1\neq
v_2$, the KS potential acquires a shelf structure in addition to the peak.  The
shelf is a direct result of the necessary alignment of the KS energy levels (the
ionization potentials) in the two fragments \cite{AB1985,M2005}. It is already
not so surprising to see that the KS potential again approaches the analytic
asymptotic form with increasing distance. As in the analytic calculation, the
exact position of the peak and the shelf depends slightly on the distance
between the two wells always being closer to the deeper well. While in the
symmetric case, see Fig.~\ref{fig:symmwall}, the dissociation limit is reached
at a distance of around 8~a.u. in the asymmetric case around 11~a.u. are
necessary. In both cases the numerical results agree perfectly with the
analytical expression. The larger distance, necessary in the asymmetric case, is
a result of the shallower right potential well in that case. 

Unfortunately, the analytical result of the shelf returning to zero can not be
verified numerically for the systems at hand. The position $x'_0$,
Eq.~(\ref{xzeroprime}), is so far away from the actual potential wells that the
density is numerically zero. There is, however, no doubt that the shelf returns
to zero exactly as predicted by the analytic formula. 

\subsection{Generalizations to three-dimensional and many-electron systems}

The general argumentation used in the previous subsection to derive the exact KS
potential in the dissociation limit is not restricted to 1D systems. The general
physical condition for dissociation is that the density is given by the sum of 
the densities of the independent fragments, Eq.~(\ref{n-diss}), because the
inter-fragment tunneling is destroyed by Coulomb correlations. The inversion
formula of Eq.~(\ref{KSpot}) is also valid for any two-particle system
independently of dimension. Therefore, an elementary 3D generalization of 
Eq.~(\ref{Vs1}) takes the form
\begin{equation}
 \label{Vs3}
\Delta v_s(\v r) = 
\frac{\left[\sqrt{n_1}\nabla\sqrt{n_2} - \sqrt{n_2}\nabla\sqrt{n_1}\right]^2}
{2(n_1 + n_2)^2} +
\frac{I_1n_1 + I_2n_2}{n_1 + n_2} - I,
\end{equation}
where all notations are the same as in Eq.~(\ref{Vs1}). Using this formula we 
can recover the {\em exact} limiting functional form of the KS potential for 
any two-particle object dissociating into two one-particle fragments. The only 
required input is the long-range asymptotics of the independent fragments, 
which is mainly determined by their ionization potentials. It is important to 
emphasize that the pre-exponential factors give only weak logarithmic 
corrections to the position of the wall and the shelf.

\begin{figure}
\includegraphics[angle=-90,width=0.48\textwidth]{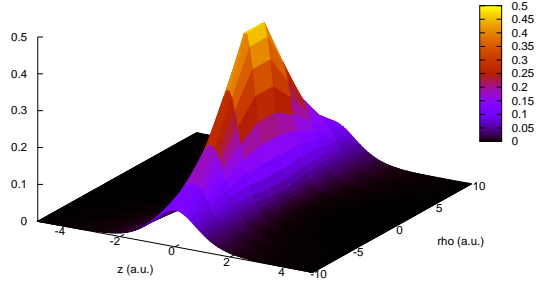}
\caption{\label{hydrogen} Kohn-Sham potential of the hydrogen molecule in the
dissociation limit.}
\end{figure}

As an illustration, we present the exact KS potential that controls the 
dissociation limit of the H$_2$ molecule. The final results obtained by 
inserting the ionization potential of the hydrogen atom, and the electronic 
densities of two independent hydrogen atoms, located at the points $\v R_1$ and 
$\v R_2$, into Eq.~(\ref{Vs3}) takes the form
\be
\label{Vs-H2}
\Delta v_s^{{\rm H}_2}(\v r)= 
\frac{1- \v r_1\v r_2/r_1r_2}{4\cosh^2(r_1 - r_2)},
\ee 
where $\v r_{1,2}= \v r - \v R_{1,2}$ are the vectors between the two protons 
and the considered point in space. The KS potential for the hydrogen molecule, 
Eq.~(\ref{Vs-H2}), is shown on Fig.~\ref{hydrogen}.
It is easy to see from Eq.~(\ref{Vs-H2}) that along the molecular axis 
$\Delta v_s^{{\rm H}_2}(\rho=0,z)$ is exactly of the 1D form 
Eq.~(\ref{Vs-delta}), while in the perpendicular direction it has a Lorentzian 
shape, 
\be
\Delta v_s^{{\rm H}_2}(\rho,z=0)=\frac{1}{2}
\frac{1}{\left(\frac{2\rho}{d}\right)^2+1},
\ee
with the width increasing at increasing distance between the two hydrogen atoms.

Similarly, we can obtain an explicit form of the exact KS potential for any
two-electron system in the strongly correlated dissociation limit. Moreover, one
can argue that the general formula (\ref{Vs3}) remains valid also for many
electron systems in those cases where the separate fragments have a single
electron in the highest occupied KS orbital. Indeed, in this case the asymptotic
behavior of the density away from the atoms is completely determined by the two
KS particles in the highest occupied KS molecular orbital (KS HOMO), while the
rest of the electrons effectively contribute to the rigid atomic cores.
Therefore,  the asymptotic form of the KS potential can be obtained by inverting
only one KS equation, namely for the KS HOMO, and, hence, the two-particle
formula (\ref{KSpot}) remains asymptotically valid.

\section{Conclusions}

In conclusion, we have presented a recipe to calculate the exact KS potential of
systems in their dissociation limit. The main ingredient is the ionization
potential of the dissociated fragments, a quantity that is readily available
from spectroscopic data. We have presented the explicit results for a
one-dimensional model system and the hydrogen molecule. It is shown that the
functional form of the potential is independent of the specific system and the 
details of the interaction as long as the latter is repulsive and sufficiently 
strong. For the 1D model system the numerical results approach the analytical
one as the distance between the two fragments is increased. Hence, they confirm
our analytical result perfectly for both a symmetric and an asymmetric system.
Our results not only pose a strong constraint for the development of
exchange-correlation functionals but also introduce an alternative way to look
at the electron localization in strongly correlated systems. How to incorporate
those effects in a density-functional treatment remains a challenge. It is
especially intriguing to explore implications of our universal results for the
quantum transport in the regime of Coulomb blockade. It is natural to expect
that the potential wall in the KS potential should modify the tunneling
probability when the transport is described in terms of KS DFT.

\acknowledgements
We acknowledge funding by the Spanish MEC (FIS2007-65702-C02-01), "Grupos
Consolidados UPV/EHU del Gobierno Vasco" (IT-319-07), and the European Community
through e-I3 ETSF project (Grant Agreement: 211956).


\begin{thebibliography}{18}

\bibitem{HK1964}
P.~Hohenberg and W.~Kohn, Phys. Rev. \textbf{136}, B864 (1964).

\bibitem{KS1965}
W.~Kohn and L.J.~Sham, Phys. Rev. \textbf{140}, A1133 (1965).

\bibitem{MR2009}
M. A. L. Marques and A. Rubio (Eds.), Phys.Chem. Chem. Phys. \textbf{11}

\bibitem{CCJ2009}
M. Casida, H. Chermette and D. Jacquemin (Eds.),  
Journal of Molecular Structure: THEOCHEM, Special issue to appear 

\bibitem{FNGB2005}
M.~Fuchs, Y.-M.~Niquet, X.~Gonze, and K.~Burke, J. Chem. Phys. 
\textbf{122}, 094116 (2005).

\bibitem{D2006}
J.~Dobson, \emph{Dispersion (Van Der Waals) Forces and TDDFT}, in 
Time-dependent density functional theory (Springer, Berlin Heidelberg, 2006),
pp. 443--462.

\bibitem{AAL1997}
V.I.~Anisimov, F.~Aryasetiawan, and A.I.~Lichtenstein, J. Phys. C \textbf{9},
767 (1997).

\bibitem{DMFT}
A. Georges, G. Kotliar, W. Krauth and M. Rozenberg, Rev. Mod. Phys.
\textbf{68}, 13 (1996), G. Kotliar and D. Vollhardt, Physics Today \textbf{57}, 53 
(2004), G. Kotliar, S.Y. Savrasov and K. Haule et al
Rev. Mod. Phys. \textbf{78}, 865 (2006)

\bibitem{AB1985}
C.O. Almbladh and U. von Barth, in \emph{Density Functional Methods in Physics} 
  (Plenum Press New York, 1985), pp. 209--231., 
J.~P. Perdew, in: \emph{Density Functional Methods in Physics} 
  (Plenum Press New York, 1985), pp. 265--308.

\bibitem{BBS1989}
M.A.~Buijse, E.J.~Baerends, and J.G.~Snijders, Phys. Rev. A \textbf{40}, 4190
(1989)

\bibitem{LB1994}
R.~van Leeuwen and E.~J. Baerends, Phys. Rev. A \textbf{49}, 2421 (1994).

\bibitem{GRB1995}
O.~V. Gritsenko, R.~van Leeuwen, and E.~J. Baerends, Phys. Rev. A
 \textbf{52}, 1870 (1995).

\bibitem{GB1996}
O.~V. Gritsenko and E.~J. Baerends, Phys. Rev. A \textbf{54}, 1957 (1996).

\bibitem{GB1997}
O.~V. Gritsenko and E.~J. Baerends, Theor. Chem. Acc. \textbf{96}, 44 (1997).

\bibitem{TMM2009}
D.G. Tempel, T.J. Mart\'inez, and N.T. Maitra, 
J. Chem. Theory and Computation, \textbf{5}, 770 (2009).

\bibitem{Girardeau1960}
M. Girardeau, J. Math. Phys. \textbf{1}, 516 (1960).

\bibitem{YukGir2005}
 V. I. Yukalov and M. D. Girardeau, Laser Phys. Lett. \textbf{2}, 375 (2005).

\bibitem{CL2006}
C.~C. Chiril$\mathrm{\breve{a}}$ and M.~Lein, Phys. Rev. A \textbf{74}, 051401R 
(2006), M.~Lein, Phys. Rev. A \textbf{72}, 053816 (2005), 
A.~Kenfack and J.~Rost, J. Chem. Phys. \textbf{123}, 204322 (2005).

\bibitem{LL1977}
L.D. Landau and E.M. Lifschitz, \emph{Quantum Mechanics} 
(Butterworth-Heinemann, 1977).

\bibitem{MCBR2003}
M. A. L. Marques, A. Castro, G. F. Bertsch, and A. Rubio, Comp. Phys. Comm. 
\textbf{151}, 60 (2003), 
A. Castro, H. Appel, M. Oliveira, C.A. Rozzi, X. Andrade, F. Lorenzen,
M. A. L. Marques, E. K. U. Gross, and A. Rubio, Phys. Stat. Sol. (b)
\textbf{243}, 2465 (2006).

\bibitem{M2005}
N.~Maitra, J. Chem. Phys. \textbf{122}, 234104 (2005).



\end{thebibliography}
\end{document}